\documentclass[runningheads]{llncs}
\usepackage[T1]{fontenc}
\usepackage{hyperref} 
\usepackage{xcolor}
\usepackage{array}
\usepackage{rotating} 
\usepackage{multirow}
\usepackage{adjustbox}
\usepackage{graphicx}
\begin{document}
\title{A RAG-Based Question-Answering Solution for Cyber-Attack Investigation and Attribution}
\titlerunning{QA Solution for Cyber-Attack Investigation and Attribution}

\author{Sampath Rajapaksha\thanks{These authors contributed equally to this work. Thus, the alphabetical order is applied.}\orcidID{0000-0001-7772-3774} 
\and
Ruby Rani$^\star$\orcidID{0000-0003-1257-8478}
\and
Erisa Karafili\orcidID{0000-0002-8250-4389}}
\authorrunning{Rajapaksha, Rani, and Karafili}

\institute{University of Southampton\\ University Road, Southampton SO17 1BJ, UK \\
\email{\{srwg1m24, r.rani, e.karafili\}@soton.ac.uk}
}

\maketitle             %
\begin{abstract}
In the constantly evolving field of cybersecurity, it is imperative for analysts to stay abreast of the latest attack trends and pertinent information that aids in the investigation and attribution of cyber-attacks. In this work, we introduce the first question-answering (QA) model and its application that provides information to the cybersecurity experts about cyber-attacks investigations and attribution. Our QA model is based on Retrieval Augmented Generation (RAG) techniques together with a Large Language Model (LLM) and provides answers to the users' queries based on either our knowledge base (KB) that contains curated information about cyber-attacks investigations and attribution or on outside resources provided by the users. We have tested and evaluated our QA model with various types of questions, including KB-based, metadata-based, specific documents from the KB, and external sources-based questions. We compared the answers for KB-based questions with those from OpenAI's GPT-3.5 and the latest GPT-4o LLMs. Our proposed QA model outperforms OpenAI's GPT models by providing the source of the answers and overcoming the hallucination limitations of the GPT models, which is critical for cyber-attack investigation and attribution. Additionally, our analysis showed that when the RAG QA model is given few-shot examples rather than zero-shot instructions, it generates better answers compared to cases where no examples are supplied in addition to the query.

\end{abstract}

\section{Introduction}
Investigating and attributing cyber-attacks are crucial processes, as they allow the implementation of efficient countermeasures. 
Cyber-attack attribution, which involves identifying the attacker responsible for a cyber-attack, is an essential process that enables experts to implement attacker-oriented countermeasures and pursue legal actions. There are three levels of attribution. The first level involves understanding the tools, tactics, techniques, and procedures (TTPs) used by the attacker. The second level identifies the geographical location of the attack. The third and most critical level determines the individual or organization behind the attack~\cite{irshad2023cyber}.
Details about cyber-attacks are published in various formats, such as text and PDF. Cybersecurity professionals often need to sift through vast amounts of these textual data to stay ahead of evolving threats~\cite{karafili2018formal,ferrag2024generative}.
The lack of a standard format for these reports makes extracting meaningful information from cyber-threat intelligence (CTI) challenging~\cite{irshad2023cyber}. Consequently, cyber-attack attribution is complex and resource-intensive~\cite{Karafili2018,Karafili2020}, often requiring manual effort from cybersecurity experts~\cite{perry2019no}. Despite its complexity, this process is essential for responding to incidents, formulating cybersecurity policies, and mitigating future threats. LLMs have revolutionised Natural Language Processing (NLP), significantly enhancing automated text generation and enabling human-like interactions. Consequently, LLMs such as GPT-3.5, GPT-4 are increasingly utilised in various domains, including cyber threat intelligence~\cite{tihanyi2024cybermetric}. Recent studies have demonstrated the feasibility of using LLMs to extract cybersecurity knowledge from large unstructured datasets~\cite{tihanyi2024cybermetric,zhang2024llms,fayyazi2023advancing,liu2023secqa}.
However, a major drawback of LLMs is their tendency to hallucinate, producing responses that are misleading or entirely fabricated. Another significant issue is the outdated knowledge of LLMs~\cite{gao2023retrieval}.
These limitations were evident in the QA task of a cybersecurity dataset in~\cite{liu2023secqa}. 
Fine-tuning with up-to-date data can be costly and still prone to hallucination~\cite{fayyazi2023advancing}. 
Retrieval Augmented Generation (RAG)~\cite{lewis2020retrieval} can address these limitations of LLMs. However, a dataset containing reliable information is necessary to aid in cyber-attack attribution. 

Given the aforementioned challenges such as the complexity of cyber-attack attribution due to the vast amount of online reports and the limitations of LLMs in providing up-to-date cybersecurity knowledge, this work introduces, to the best of our knowledge, the first Question-Answering (QA) model leveraging RAG techniques to enhance the efficiency and accuracy of cyber-attack investigation and attribution.
In particular, our RAG model uses as its KB the
AttackER dataset~\cite{deka2024}, created using text extracted from various reliable online reports and blogs focused on cyber-attack attribution and investigation. 
Our solution will provide the users (we expect them to be security and forensics analysts) information that will help during the cyber-attack investigation and attribution. The information is provided in the form of answers to the users' questions, based on past and current attacks investigation and attribution, as well as countermeasures and best practices. Our RAG-based QA model allows cybersecurity professionals and organizations to make informed decisions about cyber-attacks. 

To evaluate the model, a separate QA pairs dataset was specifically developed with a focus on the cyber-attack investigation and attribution. The model responses were evaluated using several metrics, including faithfulness, answer
relevancy, context precision, context recall, context entity recall, context relevancy, answer similarity,
and answer correctness. The 85\% and 91\% Answer relevancy score for zero-shot and few-shot respectively implies that our RAG QA model follows the instructions carefully. Similarly, a high Answer similarity score in both zero-shot and few-shot indicates a high resemblance in the generated answer and ground truth. We compared the answers for KB-based questions with those from OpenAI's GPT-3.5 and the latest GPT-4o. Our QA model outperforms OpenAI's GPT models by providing the source of the answers and overcoming the hallucination limitations of the GPT models. 
The main contributions of this paper can be summarised as follows.

\begin{itemize}
    \item Specialized KB and QA pair dataset for developing and evaluating RAG models in cyber-Attack investigation and attribution\footnote{\url{https://github.com/sampathrajapaksha/RAG-based-QA.git}}. 
        \item RAG-Based QA model to provide accurate answers for various types of questions. This model outperform OpenAI's GPT models by providing sources for its answers, thereby enhancing reliability and minimizing LLM hallucination.
    \item Chat interface with knowledge-based, private repository-based and web-based question-answering\footnote{\url{https://huggingface.co/spaces/rubypnchl/Question_Answer_Engine}}.
    \item Evaluated and analysed the performance of the RAG-based QA model for generated questions on several significant metrics to measure the reliability of generated answer, context retrieval, and hallucination.
\end{itemize}
In Sections~\ref{sec:bg} and \ref{sec:rw} we provide the background and related work. We introduce our methodology in Section~\ref{sec:methodology}. The implementation of our QA model is shown in Section~\ref{sec:implementation}, while its evaluation is provided in Section~\ref{sec:discussion}. We conclude in Section~\ref{sec:conclusion}.

\section{Background}\label{sec:bg}
This section provides an introduction to LLMs and RAG which are utilized in this work to develop the question-answering model. 

\subsection{Large Language Models (LLMs)}
Language models (LMs) are computational models capable of understanding and generating human languages. These models possess the transformative ability to predict the likelihood of word sequences or generate new text given a specific input~\cite{chang2023survey}. LMs can be categorised into four main stages of development: Statistical Language Models (SLM), Neural Language Models (NLM), Pre-trained Language Models (PLM), and Large Language Models (LLM)~\cite{zhao2023survey}. SLMs construct word prediction models based on the Markov assumption, while NLMs employ neural networks like multi-layer perceptrons (MLP) and recurrent neural networks (RNNs) to estimate the probability of word sequences.

PLMs are LMs based on the transformer architecture, incorporating a self-attention mechanism. Bidirectional Encoder Representations from Transformers (BERT)\cite{devlin-etal-2019-bert} was introduced by pre-training bidirectional language models on large, unlabelled corpora with a specific pre-training task\cite{zhao2023survey}. BERT utilises the masked language model (MLM), where some tokens in the input are randomly masked, and the model predicts the original vocabulary ID of the masked word based on contextual information alone~\cite{devlin-etal-2019-bert}. Additionally, BERT incorporates a next-sentence prediction task to jointly pre-train representations of text pairs. This design allows BERT to be fine-tuned for various NLP tasks, achieving state-of-the-art performance. Scaling up the model and training data improves its effectiveness in downstream tasks like question-answering and language inference. These larger PLMs are often referred to as LLMs~\cite{shanahan2024talking}. An innovative application of LLMs is ChatGPT, known for its human-level conversational capabilities~\cite{zhao2023survey}.
This utilizes the close source LLMs such as GPT-3.5, GPT-4 and the latest model GPT-4o developed by OpenAI. Open-source LLM include models such as Llama and Mistral by Mistral AI\footnote{\url{https://huggingface.co/spaces/open-llm-leaderboard/open_llm_leaderboard}}. 

\subsection{Retrieval Augmented Generation}
To address the issues of hallucination in LLMs and their limited access to up-to-date knowledge, the Retrieval-Augmented Generation (RAG) technique was introduced~\cite{lewis2020retrieval}. RAG addresses these issues by retrieving and integrating relevant context from external data sources not included in LLM training data, thereby enhancing the accuracy of responses. This approach combines pre-trained parametric memory, represented by a LLM model, with non-parametric memory in the form of a vector database constructed from a knowledge base. Semantic sentence embedding models, such as Sentence-BERT (SBERT), are utilised to create semantically meaningful sentence embedding of knowledge base sentences which are then stored in the vector database. Given the LLM's constraint on text length processing, it is essential to chunk the knowledge base texts into smaller segments for efficient retrieval of relevant context. A pre-trained neural retriever accesses the vector database to retrieve the most similar information based on the query. The query itself is transformed into its vector representation using the same embedding model, and similarity metrics like cosine similarity are employed to retrieve the top-k chunks relevant to the query. These chunks are then input into the generator (LLM), along with the query, to produce more accurate and contextually relevant responses.

\section{Related Work}\label{sec:rw}
Question-answering (QA) systems enable the answering of questions posed by humans in natural language, representing an advanced form of information retrieval (IR)~\cite{soares2020literature}. Early QA systems relied on techniques such as keywords, bag-of-words, Boolean logic, and regular expressions. However, these methods often struggled to produce accurate results due to their limited understanding of the user's query context and content meaning~\cite{arbaaeen2020natural}. Transformer models have significantly improved QA systems, providing human-level responses. These include encoder-based pre-trained models like BERT, ALBERT, and RoBERTa; encoder-decoder models such as BART and T5; and decoder-based LLMs like GPT, GPT-2, GPT-3, and GPT-4~\cite{nassiri2023transformer}. ChatGPT has largely replaced traditional knowledge-based QA models~\cite{tan2023can}.  

With the introduction of Retrieval-Augmented Generation (RAG) to address the limitations of LLMs, RAG has emerged as an innovative application in NLP tasks such as domain-specific question-answering (QA)~\cite{yu2024evaluation}. RAG is used across various domains, including medicine, finance, and legal fields, for QA tasks~\cite{xiong2024benchmarking,yepes2024financial,golatkar2024cpr}. To evaluate the performance of LLMs in computer security, a QA dataset with around 200 questions was introduced in~\cite{liu2023secqa}. This work generated multiple-choice questions using GPT-4 from the book ``Computer Systems Security: Planning for Success'', but it lacked human comparison and validation. 
Experiments using various LLMs revealed significant performance variations and highlighted their limitations in answering complex computer security questions. This underscores the need for domain-specific enhancements using specialized datasets and models. A benchmark dataset called CyberMetric, based on RAG for evaluating LLMs' cybersecurity knowledge, was introduced in~\cite{tihanyi2024cybermetric}. The questions were generated from cybersecurity guidelines, standards, books, and research papers. GPT-3.5 Turbo with RAG was used to generate these questions, with some randomly validated by human experts to remove questions with multiple or incorrect answers. Experiments involving 25 LLMs showed that LLMs often struggle to accurately respond to questions based on the latest research, particularly when their training data is outdated. Both studies did not focus on datasets related to cyber-attack attribution. These studies showed the limitations of LLMs for domain specific QA tasks.

In~\cite{fayyazi2023advancing}, the authors explore the application of LLMs in cybersecurity, specifically focusing on interpreting and summarizing cyber-attack Tactics, Techniques, and Procedures (TTPs) from the MITRE ATT\&CK framework. The study compares encoder-only models like RoBERTa with decoder-only models like GPT-3.5 for TTP analysis and introduces RAG to enhance the performance of decoder-only models. Using data from the MITRE ATT\&CK framework, the study finds that RAG significantly improves the interpretation of TTPs by providing relevant context, showcasing the potential of LLMs in threat intelligence. The RAG approach outperformed both direct decoder-only LLMs and encoder-only LLMs with supervised fine-tuning. 

These studies underscore the applicability of LLMs in cybersecurity tasks and emphasize the limitations and the importance of having cybersecurity domain-specific LLMs. However, none of the existing works specifically focus on cyber-attack attribution using a specific knowledge base (dataset). Therefore, our work stands out as distinct and unique in its approach and objectives.

\section{Methodology}\label{sec:methodology}
\begin{figure}[h]
    \centering    \includegraphics[width=0.8\textwidth]{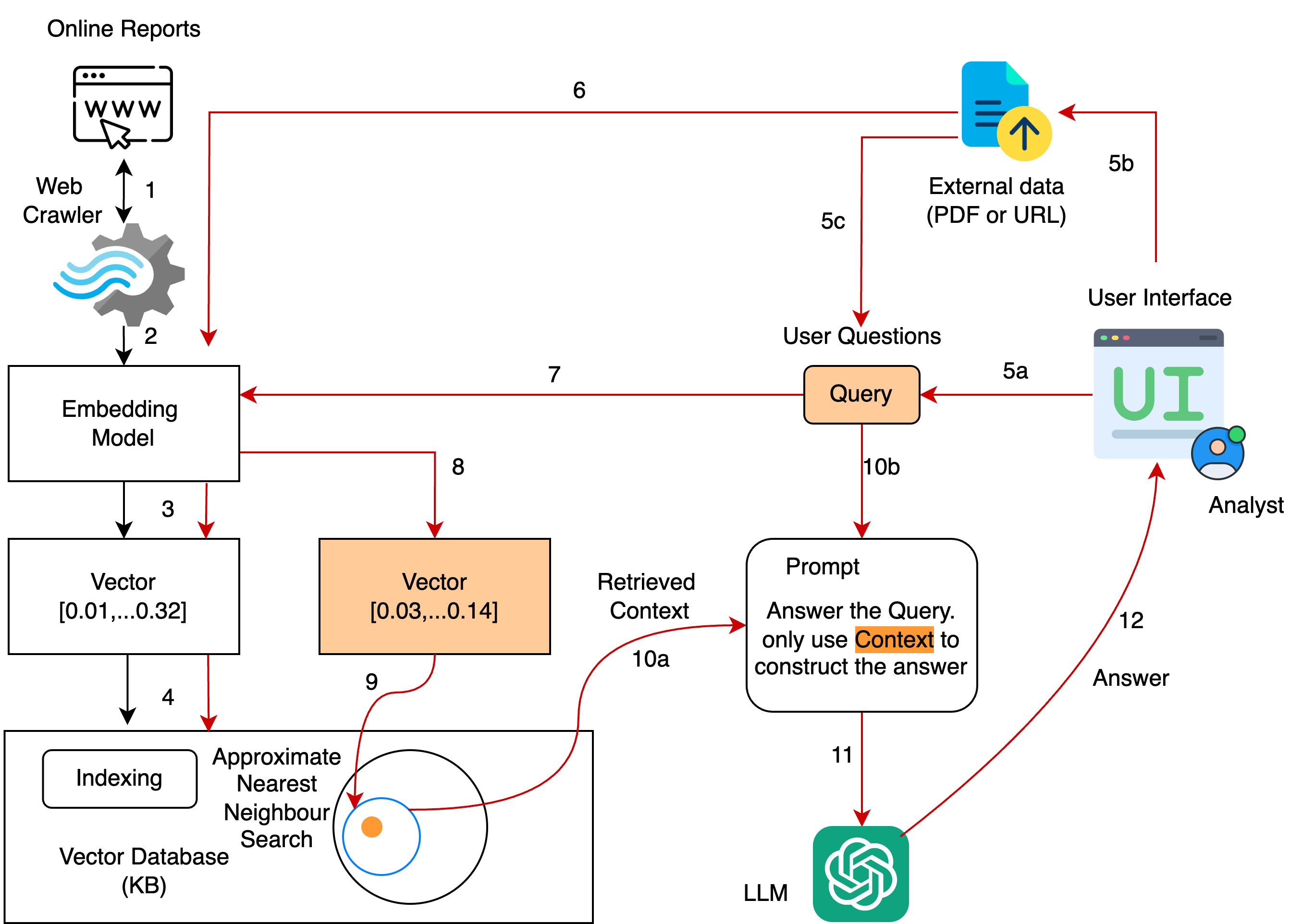}
    \caption{RAG-based QA Model Architecture}
    \label{model_architecture}
\end{figure}

In this section, we provide details about the methodology used in our question-answering model. As shown in Figure~\ref{model_architecture}, for a given user query, our model provides the answer by extracting the most similar context from the cyber-attack investigation and attribution dataset along with the source of the answer. To obtain the most similar context for the query, the approximate nearest neighbor search is performed on the vector database, and the retrieved context is used as the input prompt for the LLM along with the user query. This approach enables the LLM to provide accurate answers without relying on its outdated knowledge.
We organised the methodology section into three main parts: cyber-attack investigation and attribution knowledge base,
question-answer pair generation, and the RAG-based QA model. 

\subsection{Cyber-Attack Investigation and Attribution Knowledge Base}
Our RAG-based QA model retrieves information from the KB extracted from the AttackER dataset~\cite{deka2024}. This dataset was generated from automatically collected documents from various reliable online sources using a web crawler. These documents include reports, articles, and blogs about previous attacks written by cybersecurity experts, where past cyber-attacks were analyzed and attributed when possible. AttackER includes only documents\footnote{We ensured that the authors of the documents were reliable cybersecurity experts.} that an analyst would find valuable for gathering information. Three cybersecurity experts performed a manual analysis of the collected documents, ensuring the content was suitable and covered different types of attacks.
The main objective of the AttackER dataset was to extract information related to attack investigation and attribution from cybersecurity text to employ named entity recognition (NER). However, since the aim of our RAG application is to answer a wide variety of questions that assist in the cyber-attack investigation and attribution process, only the fields useful for this purpose were selected from the AttackER dataset, removing the annotated entities. Consequently, our knowledge based includes details such as the website name, website URL, and the extracted text. There are 202 lengthy various forms of online texts including reports, blogs, articles,etc. that include detailed descriptions of attacks, which can be queried to get answers to various questions.
The major sources of this dataset include blogs and reports from Mandiant, Malwarebytes, Talos Intelligence, Mitre, Securelist and Trendmicro. A full list is provided in Appendix~\ref{appendix}.

\subsection{Questions and Answer Pair Generation}
In this subsection, we provide details of the generation of the question-answer pairs derived from extensive online reports in the cybersecurity domain. By leveraging question-answer generation, we facilitate rapid access to crucial information, thus, supporting analysts in their investigative processes. Our research enhances the efficiency of cybersecurity investigations, by keeping the analyst up-to-date with the latest attack trends and pertinent information that aids in the investigation and attribution of cyber-attacks.

In this paper, we aim to solve the problem of answer generation for factual, contrastive, and inferential questions.
We manually generated questions with the help of two senior cybersecurity experts from the collected online reports available in the cybersecurity domain. The questions are in two different formats: the first is based on the category of questions (factual, contrastive, and inferential), while the second is based on the mode of query i.e whether the user is interested in asking the question from the entire dataset, based on provided metadata, from a specific document, or from an external source not available in our dataset. 
Our knowledge base (KB), derived from the AttackER dataset, was used to generate the questions. The number of questions generated is 280, divided between factual, contrastive, and inferential questions\footnote{Our methodology allows the usage of opinion-based questions. We decided not to include this type of question given that usually, they are not interesting for the cybersecurity experts, as well as the consistent opinions gathered inside AttackER.}. To keep the evaluation process unbiased, we carefully selected a total of 170 questions for all the categories.  
 The corresponding answers can be generated using specified query modes: querying the entire dataset, utilizing metadata details, providing specific documents, or generating answers for questions derived from external sources not present in our curated collection. We rigorously evaluated the model to ensure high performance and accuracy in generating contextually relevant answers (see Section~\ref{sec:discussion}).

\subsection{RAG-based QA Model}
Our methodology uses the RAG model architecture shown in Figure~\ref{model_architecture}. 
The red lines in the figure illustrate the processes involved during inference, while the black line indicates the initial KB creation. This architecture comprises of interconnected components, categorised primarily into: retriever and generator. The retriever relies on the knowledge base derived from the AttackER dataset
(steps 1 and 2 in the model architecture). The core component of the retriever is the vector database, which stores the precomputed vectors of the indexed knowledge base. To convert the raw texts of the knowledge base into dense vector representations suitable for storage, a sentence embedding model is employed. Sentence embedding is more appropriate for capturing the meaning and context of a sentence compared to word embedding. These embeddings are indexed in the vector database to facilitate efficient similarity search. 
The embedding and data storage processes in the vector database are illustrated in steps 3 and 4 of Figure~\ref{model_architecture}.
Once the initial KB is set up based on the AttackER dataset, the model offers two query options: querying the existing KB or querying external data by uploading a PDF file or providing a web URL. For external sources, embeddings are created and added to the knowledge base, extending it with new data (steps 5b and 6). When an analyst provides a query (question), represented in steps 5a, 5c, and 7, it undergoes transformation into its vector representation using the same embedding model (step 8). The approximate nearest neighbour (ANN) algorithm then searches the vector database (step 9), measuring similarity with the query vector using cosine similarity. This process retrieves the top-k most relevant documents (context) related to the query. Once the retriever identifies the relevant documents, they are passed to the generator model along with the query (steps 10a and 10b). The generator (a pretrained LLM), is tasked with generating answers based solely on the provided context (step 11), mitigating the risk of hallucination typical in LLMs. The generated answer is then presented in the user interface (step 12) as shown in the architecture diagram.

Furthermore, our RAG QA model generates two types of responses for each input query. It generates an answer without any examples provided, which we call zero-shot scenarios, where the model leverages its pre-existing knowledge 
and its understanding of patterns and relationships to tackle new situations. In the second type of response, called few-shot scenarios, the model improves its answers by learning from a few provided examples. The model quickly learns and performs new tasks effectively with just a little bit of guidance, using its prior knowledge to learn efficiently from limited data.

\section{RAG-based QA Model Implementation 
}\label{sec:implementation}
In this section, we provide some technical details on our RAG-based QA model implementation tool and an example. 
\subsection{QA App Deployment}
We decided to deploy the RAG model on a public server to make the our QA solution accessible and interactive. Specifically, we used the Huggingface's platform and Streamlit to create a user-friendly web application. We provide below the step-by-step deployment process. 

\begin{enumerate}
    \item \textbf{Preparation of the Streamlit Application:}
    We wrote the application script
    using Streamlit (an open-source app framework). This script handles user inputs, processes the data, generates responses using the RAG-based QA model, and displays the results. The script also includes functionalities for fetching and processing URLs, loading data asynchronously, initializing embeddings, and generating answers using open-source Large Language Models (LLMs). 

    \item \textbf{Setting Up the Environment:} We set up the Huggingface environment, obtain an API token, and install necessary libraries (e.g., Streamlit, aiohtpp, pandas, etc.).
    \item \textbf{Initializing the Huggingface Model:}
    We configured the Huggingface model in the script by setting the environment variable for the Huggingface API token. We initialized the chosen model (e.g., Mistral-7B) using the \texttt{HuggingFaceHub} class from the LangChain library.

    \item \textbf{Loading and Processing Data:}
    We implemented functions to load and process documents from various sources asynchronously. This ensures that the application can handle multiple data inputs efficiently. We use FAISS for embedding and vector storage, facilitating fast and accurate retrieval of relevant documents.

    \item \textbf{Creating the Streamlit Interface:}
    We then moved on designing the user interface in Streamlit, allowing users to select query modes, question types, and the LLM to generate answers. We also included features for handling user queries, processing CSV files, and displaying answers and source documents.

    \item \textbf{Deploying on Huggingface:}
    \begin{enumerate}
        \item We created a repository on Huggingface and uploaded our application and the needed files. 
                \item We configured the repository settings to enable Streamlit deployment.
        \item We ensured that all required dependencies are listed. 
    \end{enumerate}

    \item \textbf{Running the Application:}
    Our RAG-based QA solution will be accessible via a Huggingface-hosted URL. Users can interact with the application to ask cybersecurity-related questions and receive answers generated by the RAG-based QA model. 
\end{enumerate}

Let us now describe some further features of our RAG-based QA app.  
Users have the flexibility to choose between single question-answer generation and batch processing via a CSV file.  For single question-answer, users can query: the entire dataset, a specific document, metadata-based, or external sources-based. The response will include the question category, context, generated answer, and sources.
 The user can use a CSV file to perform the query.  The answer including question category, context, generated answer, and sources, are saved directly in the CSV file, ensuring organized and efficient data management.
 Users can the question category: factual, contrastive, opinion, or inferential—for single question inputs. For batch processing via CSV, the model autonomously determines the question category.
 The system generates answers using both zero-shot
and few-shot learning approaches. Few-shot learning involves providing the model with
example question-answer pairs, and is used to enhance response accuracy.
 Our model provides the functionality to generate answers using open-source LLMs (e.g., Mistral-7B, Llama\footnote{\url{ https://huggingface.co/meta-llama/Llama-2-7b-chat-hf}}, Zephyr\footnote{\url{https://huggingface.co/HuggingFaceH4/zephyr-7b-alpha}}) alongside the RAG pipeline. This allows users to compare the efficacy of our RAG model against other LLMs.
To ensure reliability, our QA model is designed to avoid hallucinations. If it is unsure, it will return “Sorry, I don’t know” instead of providing potentially misleading information.

\begin{figure}[t]
    \centering
    \includegraphics[width=0.5\linewidth]{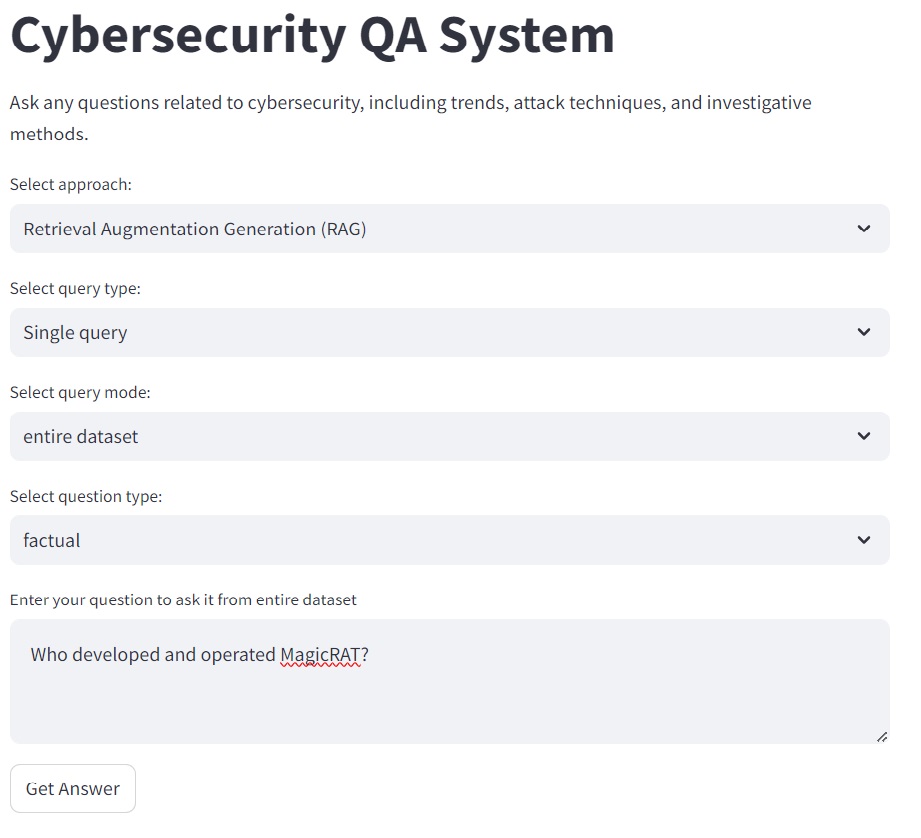}
    \caption{Query input in our QA tool}
    \label{fig:app_look_1}
\end{figure}

\begin{figure}[h]
    \centering
    \includegraphics[width=0.5\linewidth]{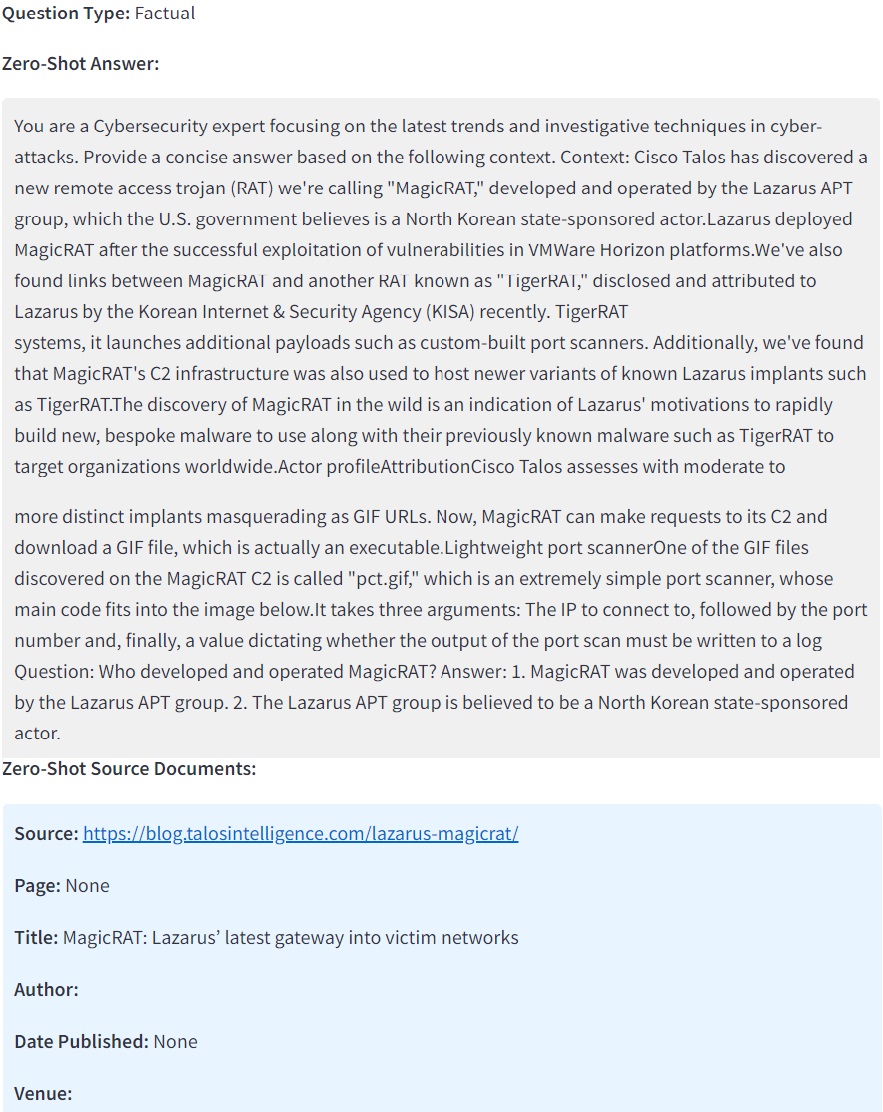}
    \caption{Generated response by our QA tool}
    \label{fig:app_qa_1}
\end{figure}

\subsection{An Example}
To illustrate the functionality and accuracy of our RAG-based QA model, we performed several single query experiments. Let us show an example thatconsider a factual query related to a specific cybersecurity incident, the developer and operator of MagicRAT, see Figures~\ref{fig:app_look_1}.
The model successfully retrieves the most relevant documents from the dataset, generates an accurate answer, and provides the context for which the information was derived, see Figure~\ref{fig:app_qa_1}. 

\section{Results and Discussion}\label{sec:discussion}
In this section, we evaluate the performance of our RAG-based QA model using the generated question-answer pairs, focusing on both the retriever and generator components.

\subsection{RAG-based QA Model Results \& Evaluation}
In our experiments, we evaluated our RAG-based QA model 
on a set of predefined questions (related to cyber security). The responses given by our solution are compared to the ground truth answers, in order to evaluate their accuracy and relevance. The set of questions was generated manually, 30 questions for each category, asked on metadata-based, specific document-based, and external sources-based except the entire dataset for which 50 questions were generated. 
We evaluate the generated questions from three main categories: factual, contrastive, and inferential. 

\subsubsection{Metrics Used}
The results were evaluated using several metrics\footnote{Each metric is calculated based on the evaluation framework proposed in~\cite{es2023ragas}.},
i.e., faithfulness, answer relevancy, context precision, context recall, context entity recall, context relevancy, answer similarity, and answer correctness, described below. 

\begin{itemize}
    \item\textbf{Faithfulness:} This measures the factual consistency of the generated answer against the given context, scaled to a range of 0 to 1, with higher values indicating better consistency.
    \item\textbf{Answer Relevancy:} This metric assesses how pertinent the generated answer is to the given prompt, with higher scores indicating better relevancy.
    \item\textbf{Context Precision:} This evaluates whether all relevant items in the contexts are ranked higher, calculated using the question, ground truth, and contexts, with values ranging from 0 to 1.
    \item\textbf{Context Recall:} This measures the extent to which the retrieved context aligns with the annotated answer, treated as the ground truth.
    \item\textbf{Context Relevancy:} This gauges the relevancy of the retrieved context based on the question and contexts, with values between 0 and 1.
    \item\textbf{Context Entity Recall:} This measures the recall of entities present in both ground truths and contexts relative to entities in the ground truths alone.
    \item\textbf{Answer Semantic Similarity:} This evaluates the semantic resemblance between the generated answer and the ground truth, with values ranging from 0 to 1.
    \item\textbf{Answer Correctness:} This assesses the accuracy of the generated answer compared to the ground truth, with scores from 0 to 1.
\end{itemize}
\begin{table}
\centering
\scalebox{0.735}{
\begin{tabular}{|p{3.7cm}|p{3.1cm}|p{3.1cm}|p{3.1cm}|p{3.1cm}|}
\hline
\multirow{2}{*}{\textbf{Metric}} & \multicolumn{2}{|c|}{\textbf{Metadata Category}} & \multicolumn{2}{|c|}{\textbf{External Source Category}} \\
\cline{2-5}
 & \textbf{Question 1} & \textbf{Question 2} & \textbf{Question 1} & \textbf{Question 2} \\
\hline
\textbf{Question} & What is the primary method of propagation for WannaCry ransomware? & Who is the likely author of the Stuxnet malware? & What is the name of the new backdoor variant used by APT29? & Which Russian threat group is linked to APT29? \\
\hline
\textbf{Answer} & The WannaCry ransomware spreads primarily through the use of the EternalBlue exploit. & The author of Stuxnet malware is not definitively known, though it is widely believed to be a collaborative effort between the US and Israeli governments. & The new backdoor variant used by APT29 is WINELOADER. & APT29 is linked to the Russian Foreign Intelligence Service (SVR). \\
\hline
\textbf{Contexts} & How to avoid ransomware ransomware threats. & References Mandiant. (2022, May 2). UNC3524: & late February, APT29 used a new backdoor variant called WINELOADER. & we judge this activity to present a broad threat to US national security. \\
\hline
\textbf{Ground Truth} & EternalBlue & novelist Mark Russinovich & WINELOADER & Russia's Foreign Intelligence Service (SVR) \\
\hline
\textbf{Faithfulness} & 0 & 0 & 0.5 & 0.666667 \\
\hline
\textbf{Answer Relevancy} & 0.951928 & 0.981471 & 0.982653 & 0.9495 \\
\hline
\textbf{Context Precision} & 0 & 0 & 0.511111 & 0.234569 \\
\hline
\textbf{Context Recall} & 0 & 0 & 1 & 1 \\
\hline
\textbf{Context Entity Recall} & 0 & 0 & 1 & 1 \\
\hline
\textbf{Context Relevancy} & 0.166667 & 0.022222 & 0.0625 & 0.055556 \\
\hline
\textbf{Answer Similarity} & 0.746007 & 0.757172 & 0.827533 & 0.837052 \\
\hline
\textbf{Answer Correctness} & 0.186502 & 0.189293 & 0.506883 & 0.509263 \\
\hline
\end{tabular}}
\caption{Evaluation of our QA responses, with metadata category and external data source category}
\label{table:evaluation_metrics_combined}
\end{table}

\begin{table}[h]
\scalebox{0.74}{
\begin{tabular}{|p{1.25cm}|p{1.95cm}|p{1.7cm}|p{1.7cm}|p{1.7cm}|p{1.7cm}|p{1.7cm}|p{1.7cm}|p{1.95cm}|}
\hline
\textbf{RAG Prompt} & \textbf{Faithfulness } & \textbf{Answer Relevancy} & \textbf{Context Precision} & \textbf{Context Recall} & \textbf{Context Entity Recall} & \textbf{Context Relevancy} & \textbf{Answer Similarity} & \textbf{Answer Correctness} \\
\hline
Zero-Shot & 0.5637998 & 0.8592762 & 0.6491907 & 0.5544218 & 0.3314383 & 0.1247952 & 0.8537770 & 0.4080567 \\
\hline
Few-Shot & 0.6249362 & 0.9171041 & 0.6603251 & 0.5854701 & 0.3474359 & 0.1147405 & 0.829127 & 0.4356158 \\
\hline
\end{tabular}}
\caption{Average metrics score of RAG model for Zero-Shot and Few-Shot prompts on entire database}
\label{table:evaluation_metrics_zero_few_shot}
\end{table}

\subsubsection{Bulk Question Evaluation}
For a more comprehensive evaluation, we used the bulk question processing feature, where users upload a CSV file containing multiple questions. We evaluated 50 questions based on the entire dataset using the RAGs library. Each question was assessed for the generated answer, context retrieved by RAG, and the ground truth prepared by two cybersecurity experts.

Details about our evaluation for the RAG-based QA model using three sample questions are provided in Table \ref{table:evaluation_metrics} (Appendix~\ref{appendixb}). 
We noticed high faithfulness scores, particularly a perfect score for the ``MagicRAT'' developer question, indicating accurate answers based on the provided contexts. Consistently high answer relevancy scores suggest the model generates pertinent answers. The model shows high context precision, especially for detecting and blocking threats, though context recall varies, indicating occasional gaps in retrieving comprehensive information. Context entity recall scores vary, with a notable 0.75 for the ``MagicRAT'' and ``TigerRAT'' questions, showing good entity extraction in some cases. Answer similarity and correctness metrics highlight the model's ability to generate semantically similar and correct answers, though there is variability, suggesting room for improvement. Overall, the RAG model demonstrates strong performance in generating accurate and relevant answers, effectively retrieving pertinent contexts, and showing robustness in handling diverse cybersecurity queries.

In Table~\ref{table:evaluation_metrics_combined}, we evaluate the RAG model's performance across metadata and external data source categories, revealing significant insights. For metadata-based queries, the model demonstrates high answer relevancy but low faithfulness and context precision, indicating relevance without factual consistency. In contrast, external data source evaluations show improved faithfulness and context recall, with high entity recall, suggesting better performance with richer data. However, variability in context precision and moderate answer correctness across both categories highlights the need for enhanced accuracy in context retrieval and factual alignment to improve the model's reliability in cybersecurity applications.

In Table \ref{table:evaluation_metrics_zero_few_shot}, we compare Zero-Shot and Few-Shot RAG prompts across several metrics. On average, Few-Shot outperforms Zero-Shot, showing higher faithfulness (0.625 vs 0.564) and answer relevancy (0.917 vs 0.859), indicating more accurate and relevant answers. However, Few-Shot has slightly lower context precision (0.660 vs 0.649), suggesting some trade-off in retrieving precise contexts. Context recall and entity recall are higher for Few-Shot, which indicates better context retrieval. Despite lower context relevancy, Few-Shot's overall improved answer correctness (0.436 vs 0.408) demonstrates its effectiveness in generating more reliable answers, making it a preferred approach for cybersecurity experts looking for critical and contextually rich responses about attack attribution.

\begin{sidewaystable}
\centering
\scalebox{0.85}{
\begin{tabular}{|p{2.5cm}|p{2.5cm}|p{4.5cm}|p{4.5cm}|p{4.5cm}|p{4.5cm}|}
\hline
\textbf{Question} & \textbf{Ground Truth} & \textbf{RAG-based QA} & \textbf{GPT-4o Answer} & \textbf{GPT-3.5 Answer} & \textbf{Analysis} \\
\hline
When did the exploitation of CVE-2023-34362 begin?&Exploitation began on May 27th, 2023, during the US Memorial Day holiday&According to the provided context, the exploitation of CVE-2023-34362 began in July 2021, but it was not widely known until the release of the security bulletin on May 31st, 2023. The actor behind the attacks was identified as Cl0p a few days later& The exploitation of CVE-2023-34362, a SQL Injection Vulnerability in Progress Software's MOVEit Transfer solution, began on May 27, 2023. It's important to apply the available patches as soon as possible to protect your systems from this vulnerability & I don't have specific information on CVE-2023-34362 as my training only includes knowledge up until January 2022 & The RAG-based QA provids a date close to the ground truth with additional details. GPT-4o also provided the correct answer aligned with the ground truth. However, GPT-3.5 does not provide an answer due to outdated training data \\
\hline
What is the relationship between MagicRAT and TigerRAT? & MagicRAT has links to TigerRAT, another RAT attributed to Lazarus, and both were found using the same C2 infrastructure. & MagicRAT and TigerRAT are both malware families that are often found together in attacks by the Lazarus APT group. MagicRAT serves as the initial entry point for an attack, often launching additional payloads such as custom-built port scanners. TigerRAT is one of the payloads that can be delivered by MagicRAT's C2 infrastructure. & MagicRAT and TigerRAT are both malware tools attributed to the Lazarus Group. MagicRAT is a Remote Access Trojan (RAT) used for espionage, while TigerRAT is another tool used by the group, indicating a broader toolkit for various cyber-espionage activities & MagicRAT and TigerRAT are both malware variants associated with the Dark Caracal threat actor group. They share similarities in their functionality and are part of the group's toolkit for conducting cyber espionage and surveillance activities & The RAG-Based QA model accurately extracts context to provide answers consistent with the ground truth. In contrast, GPT-3.5 provides incorrect answers due to hallucination or reliance on outdated knowledge. GPT-4o, while partially correct, fails to accurately identify the relationship of the C2 infrastructure.\\
\hline
What new capability was discovered in the latest variant of TigerRAT? & The latest variant includes a "USB dump" capability for enumerating and exfiltrating files with specific extensions. & USB dump. This capability allows the implant to gather data from USB devices connected to the infected system. 2. Indicators suggest that there is a second new capability set to be introduced soon. & The latest variant of TigerRAT was discovered to have a new capability of screen capturing, allowing attackers to monitor and capture the victim's screen activities in real-time & The latest variant of TigerRAT was found to have enhanced evasion techniques, including improved anti-analysis capabilities and obfuscation methods, making it more difficult for traditional security measures to detect and analyze its activities. & The RAG-Based QA model provides answers that closely match the ground truth. In contrast, both GPT-3.5 and GPT-4o provide differing answers that lack reliability due to the absence of the answer sources.\\
\hline
\end{tabular}
}
\caption{Comparison of Ground Truth, RAG-based QA , GPT-4o, and GPT-3.5 Answers - Part 1}
\label{tab:GPT_comparison_1}
\end{sidewaystable}

\subsection{Comparison of RAG-based QA model, GPT-3.5, GPT-4o answers}
In this section, we compare the results of our RAG-based QA model with the answers from GPT-3.5 and GPT-4o.
GPT-3.5 Turbo is the language model available in the free version of the ChatGPT interface. It was trained with data up to September 2021. GPT-4o is the latest advanced multimodal flagship model available through the ChatGPT interface with limited access. 
GPT-4o was trained with data up to October 2023\footnote{\url{https://platform.openai.com/docs/models/gpt-4o}}, and it has limited access to the internet for retrieving the latest data.  
Our RAG-based QA application supports questions using metadata and external data, unlike GPT-3.5, which lacks this functionality. Therefore, we perform this comparison only for questions asked from the entire knowledge base.
To ensure a fair comparison, we use the same prompt:
\textit{`You are a Cybersecurity expert focusing on the latest trends and investigative techniques in cyber-attacks. Provide a concise answer for the below question.'} for all models.
Results for the selected set of questions are summarized in Table~\ref{tab:GPT_comparison_1}
and Table~\ref{tab:GPT_comparison_2} (in Appendix~\ref{appendixb}).
Our primary findings from the model comparison can be summarized as follows:

\begin{itemize}
    \item \textbf{Unreliable answers from GPT models:} The RAG-Based QA model provides sources for answers, making it reliable for cybersecurity analysts to check and verify their validity. However, for some questions, GPT models provide answers that do not align with the ground truth, and therefore, verifying the accuracy of these critical answers for cybersecurity decision-making is not possible.
    \item \textbf{Potential for incorrect retrieval by the RAG-Based QA model:} The accuracy of the RAG-Based QA model's answers relies on the context provided, which retrieved text chunks. When there are several similar contexts available with high similarity to the query, the RAG-Based QA model may select a context that does not contain the correct answer. For instance, in the question \textit{`Who is linked to the 'Sharpshooter' campaign?'}, it retrieves a context discussing 'Sharpshooter' but not the involved party. Even if the correct context is retrieved as the second most similar, the model selects the first context, leading to an inaccurate answer.
    \item \textbf{GPT model hallucination:}
    Hallucination was observed in both GPT models during our evaluation. However, for the questions used, GPT-3.5 exhibited more susceptibility to hallucination compared to GPT-4o.
    For example, in response to the question \textit{`When did the exploitation of CVE-2023-34362 begin?'} GPT-3.5 initially responded with: \textit{`I don't have specific information on CVE-2023-34362 as my training only includes knowledge up until January 2022.'} This response correctly acknowledges the limitation of its training data. However, upon re-running the query, GPT-3.5 provided the answer: \textit{`The exploitation of CVE-2023-34362 began in early 2023, marking the start of malicious activities targeting this specific vulnerability in cybersecurity incidents.'} This indicates hallucination, as the model generated a speculative answer beyond its training scope.
\end{itemize}

In summary, our RAG-based QA model consistently delivers more reliable answers with their respective sources, which is crucial for informed cybersecurity decision-making. In comparison, GPT-4o also provides accurate answers for most of the questions, benefiting from its access to the internet, but it is limited in the number of queries that can be processed through the ChatGPT interface per day. More importantly, these responses from GPT-4o lack transparency regarding the answer source. GPT-3.5, available freely through the ChatGPT interface, is less suitable for this task due to outdated training data and tendencies towards model hallucination.

\section{Conclusion and Future Works}\label{sec:conclusion}
Cybersecurity professionals often navigate extensive volumes of textual data in reports to stay abreast of evolving threats, where the standardised CTI reports are not enough. The cyber-attack attribution is a very complex process, consequently, it requires significant manual effort from cybersecurity experts. To address this issue, RAG can be employed with a curated dataset focused on cyber-attack investigation and attribution as the knowledge base.

In this work, we proposed the first RAG-based QA model for cyber-attacks investigation and attribution. Our model, uses as a knowledge base the AttackER dataset~\cite{deka2024}, for cyber-attack attribution and investigation, together with the Mistral LLM.
The proposed RAG-based QA model aims to assist cybersecurity professionals in locating critical information reliably and making informed decisions to facilitate the cyber-attack attribution process.
Experiment results, employing various types of questions, demonstrated that the RAG-based QA model outperforms GPT-3.5 and GPT-4o due to its reliable answers and minimal model hallucination. Few-Shot RAG prompts are better than Zero-Shot, providing more accurate and relevant responses for cyber-attack attribution. Despite its superior performance, the proposed model has limitations such as occasional retrieval of incorrect contexts leading to inaccurate answers, incomplete representation of all crucial reports in our knowledge base, lack of continuous chat functionality and latency of responses due to the use of open-source solutions. 

To address these limitations, as future work, we plan to integrate AI agents for automated web crawling to continuously update our knowledge base with the latest reliable cybersecurity reports. Additionally, we aim to enhance the RAG model with advanced features like contextual chunking, reranking, and query transformation to provide more precise answers with reduced latency. We will conduct comprehensive evaluations to further analyse and validate the proposed methods.

\section*{Acknowledgement}
Research funded by the University of Southampton on behalf of the Defence Science and Technology Laboratory (Dstl) which is an executive agency of the UK Ministry of Defence providing world class expertise and delivering cutting-edge science and technology for the benefit of the nation and allies. The research supports the Autonomous Resilient Cyber Defence (ARCD) project within the Dstl Cyber Defence Enhancement programme.

 \bibliographystyle{splncs04}
\bibliography{reference}

\begin{thebibliography}{10}
\providecommand{\url}[1]{\texttt{#1}}
\providecommand{\urlprefix}{URL }
\providecommand{\doi}[1]{https://doi.org/#1}

\bibitem{arbaaeen2020natural}
Arbaaeen, A., Shah, A.: Natural language processing based question answering
  techniques: A survey. In: 2020 IEEE 7th International Conference on
  Engineering Technologies and Applied Sciences (ICETAS). pp.~1--8. IEEE (2020)

\bibitem{chang2023survey}
Chang, Y., Wang, X., Wang, J., Wu, Y., Yang, L., Zhu, K., Chen, H., Yi, X.,
  Wang, C., Wang, Y., et~al.: A survey on evaluation of large language models.
  ACM Transactions on Intelligent Systems and Technology  (2023)

\bibitem{deka2024}
Deka, P., Rajapaksha, S., Rani, R., Almutairi, A., Karafili, E.: Attacker:
  Towards enhancing cyber-attack attribution with a named entity recognition
  dataset. arXiv:2408.05149  (2024), \url{https://arxiv.org/abs/2408.05149}

\bibitem{devlin-etal-2019-bert}
Devlin, J., Chang, M.W., Lee, K., Toutanova, K.: {BERT}: Pre-training of deep
  bidirectional transformers for language understanding. In: Burstein, J.,
  Doran, C., Solorio, T. (eds.) Proceedings of the 2019 Conference of the North
  {A}merican Chapter of the Assoc. for Comp. Ling.: Human Language
  Technologies, Volume 1. pp. 4171--4186 (2019)

\bibitem{es2023ragas}
Es, S., James, J., Espinosa-Anke, L., Schockaert, S.: Ragas: Automated
  evaluation of retrieval augmented generation. arXiv:2309.15217  (2023)

\bibitem{fayyazi2023advancing}
Fayyazi, R., Taghdimi, R., Yang, S.J.: Advancing ttp analysis: Harnessing the
  power of encoder-only and decoder-only language models with retrieval
  augmented generation. arXiv:2401.00280  (2023)

\bibitem{ferrag2024generative}
Ferrag, M.A., Alwahedi, F., Battah, A., Cherif, B., Mechri, A., Tihanyi, N.:
  Generative ai and large language models for cyber security: All insights you
  need. arXiv:2405.12750  (2024)

\bibitem{gao2023retrieval}
Gao, Y., Xiong, Y., Gao, X., Jia, K., Pan, J., Bi, Y., Dai, Y., Sun, J., Wang,
  H.: Retrieval-augmented generation for large language models: A survey.
  arXiv:2312.10997  (2023)

\bibitem{golatkar2024cpr}
Golatkar, A., Achille, A., Zancato, L., Wang, Y.X., Swaminathan, A., Soatto,
  S.: Cpr: Retrieval augmented generation for copyright protection. In:
  Proceedings of the IEEE/CVF Conference on Computer Vision and Pattern
  Recognition. pp. 12374--12384 (2024)

\bibitem{irshad2023cyber}
Irshad, E., Siddiqui, A.B.: Cyber threat attribution using unstructured reports
  in cyber threat intelligence. Egyptian Informatics Journal  \textbf{24}(1),
  43--59 (2023)

\bibitem{karafili2018formal}
Karafili, E., Cristani, M., Vigan{\`o}, L.: A formal approach to analyzing
  cyber-forensics evidence. In: ESORICS 2018, 23rd European Symposium on
  Research in Computer Security. vol. 11098, pp. 281--301 (2018)

\bibitem{Karafili2018}
Karafili, E., Wang, L., Kakas, A., Lupu, E.C.: Helping forensic analysts to
  attribute cyber-attacks: an argumentation-based reasoner. In: PRIMA 2018:
  Principles and Practice of Multi-Agent Systems. Springer International
  Publishing (2018)

\bibitem{Karafili2020}
Karafili, E., Wang, L., Lupu, E.C.: An argumentation-based reasoner to assist
  digital investigation and attribution of cyber-attacks. Forensic Science
  International: Digital Investigation  \textbf{32}(2020),  300925 (2020)

\bibitem{lewis2020retrieval}
Lewis, P., Perez, E., Piktus, A., Petroni, F., Karpukhin, V., Goyal, N.,
  K{\"u}ttler, H., Lewis, M., Yih, W.t., Rockt{\"a}schel, T., et~al.:
  Retrieval-augmented generation for knowledge-intensive nlp tasks. Advances in
  Neural Information Processing Systems  \textbf{33},  9459--9474 (2020)

\bibitem{liu2023secqa}
Liu, Z.: Secqa: A concise question-answering dataset for evaluating large
  language models in computer security. arXiv:2312.15838  (2023)

\bibitem{nassiri2023transformer}
Nassiri, K., Akhloufi, M.: Transformer models used for text-based question
  answering systems. Applied Intelligence  \textbf{53}(9),  10602--10635 (2023)

\bibitem{perry2019no}
Perry, L., Shapira, B., Puzis, R.: No-doubt: Attack attribution based on threat
  intelligence reports. In: 2019 IEEE ISI. pp. 80--85 (2019)

\bibitem{shanahan2024talking}
Shanahan, M.: Talking about large language models. Communications of the ACM
  \textbf{67}(2),  68--79 (2024)

\bibitem{soares2020literature}
Soares, M.A.C., Parreiras, F.S.: A literature review on question answering
  techniques, paradigms and systems. Journal of King Saud University-Computer
  and Information Sciences  \textbf{32}(6),  635--646 (2020)

\bibitem{tan2023can}
Tan, Y., Min, D., Li, Y., Li, W., Hu, N., Chen, Y., Qi, G.: Can chatgpt replace
  traditional kbqa models? an in-depth analysis of the question answering
  performance of the gpt llm family. In: International Semantic Web Conference.
  pp. 348--367. Springer (2023)

\bibitem{tihanyi2024cybermetric}
Tihanyi, N., Ferrag, M.A., Jain, R., Debbah, M.: Cybermetric: A benchmark
  dataset for evaluating large language models knowledge in cybersecurity.
  arXiv:2402.07688  (2024)

\bibitem{xiong2024benchmarking}
Xiong, G., Jin, Q., Lu, Z., Zhang, A.: Benchmarking retrieval-augmented
  generation for medicine. arXiv:2402.13178  (2024)

\bibitem{yepes2024financial}
Yepes, A.J., You, Y., Milczek, J., Laverde, S., Li, L.: Financial report
  chunking for effective retrieval augmented generation. arXiv:2402.05131
  (2024)

\bibitem{yu2024evaluation}
Yu, H., Gan, A., Zhang, K., Tong, S., Liu, Q., Liu, Z.: Evaluation of
  retrieval-augmented generation: A survey. arXiv:2405.07437  (2024)

\bibitem{zhang2024llms}
Zhang, J., Bu, H., Wen, H., Chen, Y., Li, L., Zhu, H.: When llms meet
  cybersecurity: A systematic literature review. arXiv:2405.03644  (2024)

\bibitem{zhao2023survey}
Zhao, W.X., Zhou, K., Li, J., Tang, T., Wang, X., Hou, Y., Min, Y., Zhang, B.,
  Zhang, J., Dong, Z., et~al.: A survey of large language models.
  arXiv:2303.18223  (2023)

\end{thebibliography}

\appendix
\section{Appendix A} \label{appendix}
Other sources from where AttackER collected data include BleepingComputer, HexaCorn, SentinelOne, CrowdStrike, Reuters, Att, Kaspersky, Webroot, Welivesecurity, Virusbulletin, Tadviser, Forumspb, Netresec, Brighttalk, Libevent, Fortinet, Microsoft, Washingtonpost, Reversemode, Viasat, Wikipedia, Wired, Cisa, Airforcemag, Businesswire, Cyberuk, Proofpoint, Fb, Withsecure, Techcrunch, Mozilla, Humansecurity, Nist, Intel471, Morphisec, Payplug, Sophos, Coretech, Stratixsystems, Crayondata, Medium, Cybergeeks, Gridinsoft, Securin, Rsisecurity, ITgovernance, DigitalGuardian, IronNet, ThreatConnect, ProtectUK, Forbes.

\section{Appendix B} \label{appendixb}
In Table~\ref{table:evaluation_metrics} we provide the evaluation for our RAG-based QA model responses on Entire Dataset, while in Table~\ref{tab:GPT_comparison_2} we provide the second part of the comparison of ground truth with our RAG-based QA model GPT-4o, and GPT-3.5 answers.

\begin{sidewaystable}
 \scalebox{0.75}{
\centering
\begin{tabular}{|p{1.5cm}|p{4.5cm}|p{3cm}|p{2.5cm}|p{2cm}|p{1.5cm}|p{1.5cm}|p{1.5cm}|p{1.5cm}|p{1.5cm}|p{1.5cm}|p{1.5cm}|}
\hline
\textbf{Question} & \textbf{Answer} & \textbf{Contexts} & \textbf{Ground Truth} & \textbf{Faithfulness} & \textbf{Answer Relevancy} & \textbf{Context Precision} & \textbf{Context Recall} & \textbf{Context Entity Recall} & \textbf{Context Relevancy} & \textbf{Answer Similarity} & \textbf{Answer Correctness} \\
\hline
Who developed and operated MagicRAT? & Answer: 1. The developer and operator of MagicRAT is the Lazarus Advanced Persistent Threat (APT) group, which is believed to be a North Korean state-sponsored actor. & [Cisco Talos has discovered a new remote access... & The Lazarus APT group, which is believed to be a North Korean state-sponsored actor. & 1 & 0.952767 & 0.844444 & 1 & 1 & 0.071429 & 0.872344 & 0.518086 \\
\hline
What is the relationship between MagicRAT and TigerRAT? & Answer: 1. MagicRAT and TigerRAT are both malware families that are often found together in attacks by the Lazarus APT group. MagicRAT serves as the initial entry point for an attack, often launching additional payloads such as custom-built port scanners. TigerRAT is one of the payloads that can be delivered by MagicRAT's C2 infrastructure. & [systems, it launches additional payloads such as custom-built port scanners. TigerRAT is one of the payloads that can be delivered by MagicRAT's C2 infrastructure.] & MagicRAT has links to TigerRAT, another RAT attributed to Lazarus, and both were found using the same C2 infrastructure. & 0.666667 & 0.967987 & 0.817857 & 1 & 0.75 & 0.25 & 0.897425 & 0.801279 \\
\hline
What are some ways to detect and block threats? & Answer: 1. Block common forms of entry: Ensure that your systems are secured against known vulnerabilities. & [can be used to identify which (if any) of the... & Using Cisco Secure Endpoint, Cisco Secure Web Appliance, and other tools. & 0 & 0.834442 & 1 & 0 & 0 & 0.384615 & 0.784633 & 0.196158 \\
\hline
\end{tabular}}
\caption{Evaluation metrics for RAG-based QA model responses on Entire Dataset based Category}
\label{table:evaluation_metrics}
\end{sidewaystable}

\begin{sidewaystable}
\centering
 \scalebox{0.83}{
\begin{tabular}{|p{2cm}|p{2cm}|p{5cm}|p{5cm}|p{5cm}|p{5cm}|}
\hline
\textbf{Question} & \textbf{Ground Truth} & \textbf{RAG-Based QA} & \textbf{GPT-4o Answer} & \textbf{GPT-3.5 Answer} & \textbf{Analysis} \\
\hline
Who developed and operated MagicRAT? & The Lazarus APT group, which is believed to be a North Korean state-sponsored actor. & The developer and operator of MagicRAT is the Lazarus Advanced Persistent Threat (APT) group, which is believed to be a North Korean state-sponsored actor. This group is also known to have developed and operated other malware, such as TigerRAT. & MagicRAT was developed and operated by the Lazarus Group, a state-sponsored hacking group believed to be associated with North Korea& MagicRAT was developed and operated by a threat actor group known as "Dark Caracal" & Both the RAG-Based QA model and GPT-4o consistently provide accurate answers that align with the ground truth. In contrast, GPT-3.5 occasionally provides incorrect answers, likely due to its reliance on outdated knowledge.\\
\hline
Who is linked to the 'Sharpshooter' campaign?&North Korea's Lazarus threat actor&Sharpshooter is linked to an unnamed threat actor. The threat actor was first detected in late 2018 and their activity has been ongoing since at least September 2017.  The threat actor launched attacks on at least 87 organizations across the globe, predominantly in the defense and critical infrastructure sectors, during the two months towards the end of last year&The 'Sharpshooter' campaign is linked to the North Korean state-sponsored hacking group known as Lazarus Group&The 'Sharpshooter' campaign is linked to the Lazarus Group, a cybercrime group associated with North Korea, known for carrying out sophisticated cyber espionage and financially motivated attacks globally&Both GPT-3.5 and GPT-4o accurately provide the correct answer aligned with the ground truth. However, the RAG-Based QA model fails to do so because it retrieved context that did not include the answer. This discrepancy arose because another web article in our knowledge base discussed 'Sharpshooter' without mentioning the party linked to it.\\
\hline
Since when were the new variants of the Rising Sun backdoor used?&Since at least 2016&Since at least 2016, the new variants of the Rising Sun backdoor have been in use&The new variants of the Rising Sun backdoor have been used since at least 2018&The new variants of the Rising Sun backdoor have been observed in use since 2018, targeting organizations primarily in Japan and South Korea, with a focus on espionage and data exfiltration&While both GPT-3.5 and GPT-4o indicate the answer is 2018, the RAG-Based QA model correctly identifies the accurate answer as 2016\\
\hline
\end{tabular}}%
\caption{Comparison of Ground Truth, RAG-based QA , GPT-4o, and GPT-3.5 Answers - Part 2}
\label{tab:GPT_comparison_2}
\end{sidewaystable}

\end{document}